\documentclass[useAMS,usegraphicx,usenatbib]{mn2e}

\usepackage{graphicx}
\usepackage{times}
\usepackage{amssymb}
\usepackage{euscript}
\usepackage{amsmath}
\usepackage{mathrsfs}

\def\chandra{{\it Chandra}}

\def\xmm{{\it XMM-Newton}}

\def\rosat{{\it ROSAT}}

\def\xeus{{\it XEUS}}

\def\s{{\rm\thinspace s}}
\def\ks{{\rm\thinspace ks}}
\def\ms{{\rm\thinspace Ms}}

\def\ctps{{\rm\thinspace ct\ s^{-1}}} 

\def\kev{{\rm\thinspace keV}}

\def\km{{\rm\thinspace km}}

\def\cm{{\rm\thinspace cm}}

\def\kpc{{\rm\thinspace kpc}}

\def\cpersperacrminsq{{\rm\thinspace counts \thinspace s^{-1} \thinspace arcmin^{-2}}}

\def \dd{{\rm d}}
\usepackage{wasysym}

\title[]{Detecting Sound-Wave-Like Surface Brightness Ripples in Cluster Cores}
\author[J. Graham, A.C. Fabian \& J.S. Sanders]{J. Graham$^1$\thanks{E-mail:jgraham@ast.cam.ac.uk}, A.C. Fabian$^1$ and J.S. Sanders$^1$\\
  \footnotesize$^1$ Institute of Astronomy, Madingley Road, Cambridge}

\begin{document}
\maketitle

\begin{abstract}
  We investigate the observational requirements for the detection of
  sound-wave-like features in galaxy cluster cores. We calculate the
  effect of projection on the observed wave amplitude, and find that
  the projection factor depends only weakly on the underlying cluster
  properties but strongly on the wavelength of the sound waves, with
  the observed amplitude being reduced by a factor $\sim 5$ for
  $5\kpc$ waves but only by a factor $\sim2$ for $25\kpc$ waves. We go
  on to estimate the time needed to detect ripples similar to those
  previously detected in the Perseus cluster in other clusters. We
  find that the detection time scales most strongly with the flux of
  the cluster and the amplitude of the ripples. By connecting the
  ripple amplitude to the heating power in the system, we estimate
  detection times for a selection of local clusters and find that
  several may have ripples detected with $\sim1\ms$ \chandra\ time.
\end{abstract}

\begin{keywords}
\end{keywords}

\section{Introduction}

The thermal history of the hot gas in galaxy clusters has been a
long-standing puzzle to astronomers. In many clusters (often referred
to as cool-core clusters), the cooling time of the hot X-ray emitting
IntraCluster Medium (ICM) drops to a fraction of the age of the
universe in the brightest central regions. The first models of these
clusters suggested that a significant amount of the gas should form a
``cooling flow'', in which gas in the cluster core cools out of the
X-ray band, reducing the pressure support on the surrounding material
and causing the outer gas to flow toward the centre. The cool gas
predicted by this model is expected to fuel star formation in the
brightest cluster galaxy. However, it has long been known that the
star formation rates in these systems are significantly lower than
these cooling flow models would predict.

With the launch of the most recent generation of X-ray satellites ---
particularly \chandra\ and \xmm\ --- it has been shown that the amount
of cool X-ray emitting gas in clusters is much smaller than predicted
by the simple cooling flow models \citep{Peterson2003}. This has led
to the widely accepted conclusion that the ICM is being heated,
reducing the cooling flow to around $10$ percent of the expected value
(see e.g. \citealt{PetersonFabian2006} or \citealt{McNamaraNulsen2007}
for reviews).

At the same time that \chandra\ and \xmm\ have confirmed that heating
of cluster cores is required, there have been substantial advances in
identifying possible heating mechanisms. Whilst there are a large
number of possible heating mechanisms, the current leading contender
is mechanical energy injection by the central black hole. Evidence of
the interaction between the ICM and the central cluster was first
noted in \rosat\ observations of the Perseus cluster
\citep{Bohringer1993}, with relativistic plasma from the AGN jets
displacing the ICM to form radio-emitting cavities. Since that time,
detection of such cavities in nearby clusters has become commonplace,
with a recent study by \citet{Dunn2006} showing evidence for X-ray
cavities in at least $70$ percent of the clusters in their sample with
short central cooling times and central temperature drops. The work
done on the ICM in the inflation of these cavities provides a
plausible source for the heating of the ICM, and several studies have
shown that the power associated with the bubbles' inflation is
sufficient to offset the cooling luminosity \citep{Birzan2004,
  Dunn2004, Rafferty2006, Dunn2005}.

Despite the promise shown by the AGN-injection model of cluster-core
heating, there are a substantial number of details that remain
unclear. One particular problem is how the energy injected by the AGN
near the cluster core is distributed over the whole cooling region in
a quasi-isotropic manner. The peaked central metal abundances in many
cool-core clusters indicate that this process must be gentle,
suggesting that it is not simply a matter of clusters having
occasional unusually large outbursts that heat the outer regions.

\begin{figure}
  \includegraphics[width=\columnwidth]{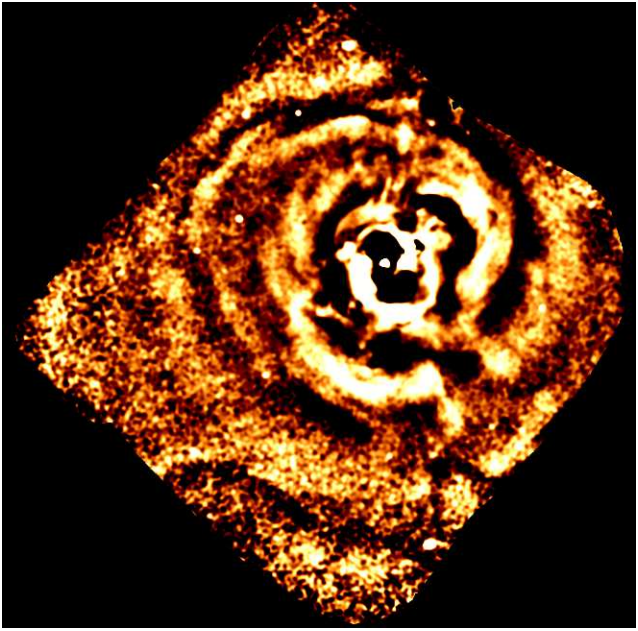}
  \caption{A $900\ks$ \chandra\ image of the Perseus cluster core,
    with a high pass filter applied to remove the underlying surface
    brightness profile and show the ripples. It is clear that the
    ripples are highly symmetric about the central source, but not
    coherent over the full $2\pi$. Reproduced from \citet{Sanders2007}}
  \label{fig:per_ripples}
\end{figure}

A possible clue to the method of energy transport away from the
cluster core was given by the unexpected detection of quasi-spherical
ripples in the surface brightness structure in the Perseus cluster
\citep{Fabian2003}, shown in Fig. \ref{fig:per_ripples} with a high
pass filter applied to remove the underlying cluster emission
(reproduced from \citealt{Sanders2007}). Ripples in surface brightness
are indicative of fluctuations in the density of the ICM and
consequentially, these features have been interpreted as sound waves
generated by the cavity inflation process. Such sound waves are
expected to be effective in distributing the AGN's mechanical energy
input over the entire cluster core \citep{McNamaraNulsen2007} and, in
the case of Perseus, it has been shown that the energy they carry is
comparable to that required to offset the cooling \cite{Sanders2007}.

Whilst cavities are common, associated sound waves have never been
detected in a cluster other than Perseus. However, the Perseus cluster
is both a factor of $\sim2$ brighter than any other cluster and has a
factor $2$ more observation time than any other cluster. In order to
understand whether sound waves do indeed play a general role in the
heating of cluster cores, however, their presence needs to be
confirmed in other systems. Our main aim in this paper is to
investigate the relationship between the intrinsic and observed
properties of sound waves in galaxy clusters and to assess the
prospects for detecting sound waves in systems other than Perseus. To
do this, we will not try to model the detailed time evolution of
ripples in cluster cores, but will assume that the
almost-monochromatic, low amplitude, ripples in Perseus are a good
template for the features expected in other clusters. By imposing
ripples derived from this template on model clusters, we will
determine how the detection time varies with the properties of the
ripples and the underlying cluster profile.

Throughout the paper we adopt a cosmology where $H_0=70\km \s^{-1} \kpc^{-1}$,
$\Omega_M = 0.3$ and $\Omega_\Lambda = 0.7$. The redshift of the
Perseus cluster is 0.018.

\section{Effect of projection on ripple amplitudes}\label{sec:projection}

Inferring information about extended sources such as galaxy clusters
is complicated by the fact that, whilst the source is intrinsically
three dimensional, our view of it is only two dimensional. In the case
of an idealised spherically symmetric cluster, the image as viewed on
the sky is related to the underlying emission profile by the Abel
integral:
\begin{equation}
  I(b) = \int^{\infty}_{b} \frac{2r\epsilon(r)}{\sqrt{r^2-b^2}},
\end{equation}
where $b$ is the projected distance on the sky, $r$ is the
3-dimensional radius, $\epsilon$ is emissivity and $I$ is the
surface brightness.

Ignoring the effect of projection, a monochromatic density wave in an
isothermal galaxy cluster with $\epsilon \propto n^{2}_{e}$ will lead to
an emissivity profile of the form
\begin{equation}
\epsilon \propto \left( \left[1+f\sin\left(\frac{2\pi}{\lambda}r\right)\right]n\right)^{2},
\end{equation}
where $f$ is the fractional amplitude of the density wave. The
corresponding amplitude of the emissivity fluctuations is then
\begin{align}
h_{\epsilon} = \frac{\left( \left[1+f\sin\left(\frac{2\pi}{\lambda}r\right)\right]n\right)^{2} -
  n^{2}}{n^{2}}\\ 
 = f^{2}\sin^{2}\left(\frac{2\pi}{\lambda}r\right) - 2f\sin\left(\frac{2\pi}{\lambda}r\right)
\end{align}
i.e. the amplitude of the ripples in emissivity is $|h_{\epsilon}|\sim 2f$.

However, as noted by \cite{Fabian2006}, the effect of projection is to
reduce the magnitude of the surface brightness ripples so that in
general $|h_{sb}|<|h_{\epsilon}|$, with troughs in the emissivity
being filled by the surrounding peaks, and vice-versa. For the case of
the Perseus cluster, \citep{Fabian2006} found that projection acts
such that $|h_{sb}| \sim f/2.5 \sim |h_{\epsilon}|/5$.  However this
result is not universal as the extent to which a trough in emissivity
is filled by the surrounding peaks depends on the relative overlap and
emission of the peaks and troughs along the line of sight. This is a
function both of the underlying surface brightness profile and the
properties of the ripples themselves.

\begin{figure}
  \includegraphics[width=\columnwidth]{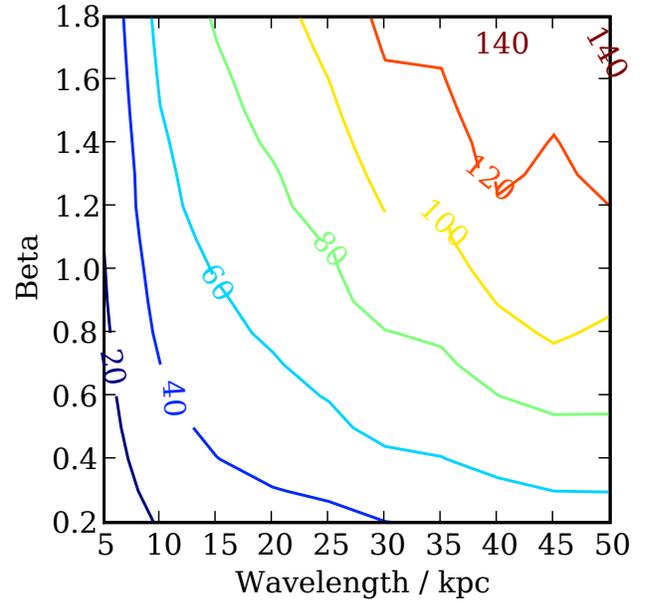}
  \caption{Amplitude of surface brightness ripple height as a
    percentage of the density ripple amplitude for a range of
    underlying density model $\beta$ parameters and density
    perturbation wavelengths and with fixed amplitude density perturbation
    $f=0.15$ and fixed density model $r_\text{core}=25\kpc$.}
  \label{fig:height_wavelength_beta}
\end{figure}

To quantify the relationship between the wavelength and cluster
properties and the projected ripple height, we have numerically
calculated the projected ripple height for monochromatic ripples in
clusters with an underlying beta-model density distribution
\citep{Cavaliere1976}:
\begin{equation}
n_{e} = \frac{n_{e,0}}{\left(1+\left(r/r_\text{c}\right)^{2}\right)^{3\beta/2}},
\end{equation}
where $r_\text{c}$ is the core radius of the cluster, $\beta$ is the
beta parameter and $n_{e,0}$ is a normalisation. The beta model
provides a reasonable fit to the density distribution in many clusters
except in the innermost region where adding a second beta-model
component often provides a better fit.

\begin{figure}
  \includegraphics[width=\columnwidth]{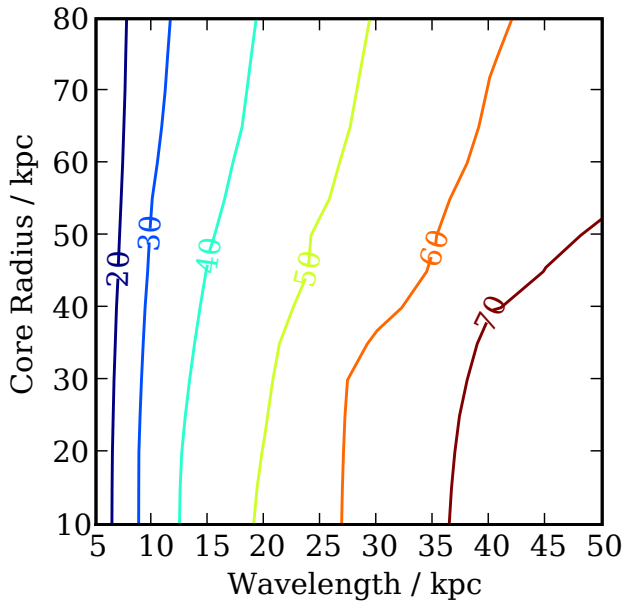}
  \caption{Amplitude of the surface brightness ripples as a percentage
    of the density ripple amplitude for a range of ripple wavelengths
    and beta-model core radii for fixed amplitude density perturbation
    $f=0.15$ and fixed $\beta=0.6$. The amplitude is almost independent of core radius for
    wavelengths below $25\kpc$.}
  \label{fig:height_wavelength_coreradius}
\end{figure}

Fig \ref{fig:height_wavelength_beta} shows the median projected ripple height
calculated for a isothermal $kT=6\kev$ cluster with an underlying
beta-model density distribution,
\begin{equation}
\frac{n_{e}}{\cm^{-3}} = \frac{4.5\times10^{-2}}{\left(1+\left(r/25\kpc\right)^{2}\right)^{3\beta/2}},
\end{equation}
with values of $\beta$ in the range $0.2-1.8$ and monochromatic density
perturbations with $f=0.15$ and wavelengths in the range $5-50\kpc$. The
use of the median height reflects the fact that not all ripples have
exactly the same height in projection, however the heights are
consistent enough that the exact method of quantifying $|h_{sb}|$ for
a given set of parameters does not significantly affect the
results. The projected height decreases with
shorter ripple wavelengths and shallower density profiles, which is
expected as these are the conditions in which adjacent peaks and troughs in
emissivity will have the most similar amplitude along the line of
sight. Fig. \ref{fig:height_wavelength_coreradius} shows the projected
height for a similar cluster but with $\beta=0.6$ and a range of
core-radii. It is apparent that the amplitude of the surface
brightness ripples is not a strong function of the core radius,
although larger core radii --- flatter emissivity profiles ---
do tend to show smaller surface brightness ripples, as expected.

Testing clusters with fixed $\beta$ and core radius, but varying
density amplitude, shows a linear variation in the projected amplitude
over the entire range of density amplitudes likely to be relevant in
clusters. This means that our calculations for a single ripple
amplitude scale in a simple way to all amplitudes.


\section{Time to Detect Ripples in Clusters}\label{sec:time}

\subsection{A Naive Approach}
The basic condition to detect a ripple is that the variation in the
number of counts in the peaks and troughs of the surface brightness
profile are distinguishable from variations due purely to
noise. Assuming a regime where Poisson noise dominates, this implies
that to detect a single peak we require an observing time such that:
\begin{equation}
(n_{\text{ripple}}-n_{\text{no ripple}})t > \sigma \sqrt{n_{\text{no ripple}}t}
\end{equation}
where $n_{\text{ripple}}$ is the average count rate in the ripple
peak, $n_{\text{no ripple}}$ is the average count rate that would be
observed in the peak without a ripple, $t$ is the time to detect the
ripple, and $\sigma$ is the number of standard deviations above the mean we
require for a detection. Therefore
\begin{equation}
t > \frac{\sigma^{2} n_{\text{no ripple}}}{(n_{\text{ripple}}-n_{\text{no ripple}})^{2}}.
\end{equation}
Taking the underlying density profile to be a beta model and the
cluster to be isothermal, the surface brightness profile will be a
projected beta model, of the form \citep{Ettori2000}:
\begin{equation}
S(r) = S_0\left(1+\left(r/r_\text{c}\right)^{2}\right)^{0.5-3\beta}.
\end{equation}

If the cluster is perfectly spherically symmetric
and ripples are sinusoidal with a wavelength $\lambda$ and constant
fractional surface brightness amplitude $h$ the detection time is:

\begin{multline}\label{eqn:t_detect_analytic}
t > \frac{\sigma^{2}}{ 2\pi S_{0}} \left (\int_{r_0}^{r_0 + \lambda/2}r\Phi(r) \dd r \right ) \\
\Bigg(\int_{r_0}^{r_0 + \lambda/2}r
    \left(1+f\sin\left(\frac{2\pi r}{\lambda}\right)\right)\Phi(r)\dd r - \\
\int_{r_0}^{r_0 + \lambda/2}r\Phi(r) \dd r \Bigg)^{-1}
\end{multline}
where
\begin{equation}
\Phi(r) = \left(1+ (\frac{r}{r_{\text{c}}})^2 \right)^{0.5-3\beta},
\end{equation}
and $S(r) = S_0\Phi(r)$ represents the projected surface brightness in counts
per second per unit area at radius $r$. $r_0$, the lower limit of
integration, is an integer number of wavelengths from the centre so
the surface brightness is integrated over a peak. Taking some
parameters appropriate for a Perseus-like cluster at $z\sim 0.018$;
$\beta=0.5$, $r_{\text{c}}=30\kpc$, $\lambda=15\kpc$, $h=0.05$,
$S_{0}=5\cpersperacrminsq$ and $\sigma=3$, we find
$t_{\text{detect}}=1.5\ks$. This is clearly short compared to the time
actually required to detect the sound waves in Perseus, which were not
observed in a $25\ks$ observation but were observed in a $200\ks$
observation. Part of this discrepancy can be explained by the fact
that our calculation assumes that the ripples are coherent over a
entire annulus of the cluster. In practice, the ripples in Perseus are
coherent over a much smaller angle. To detect a ripple over some
fraction $\chi$ of an angle requires that the integration time is
increased by a corresponding factor of $1/\chi$.

Although this method is simple and easy to apply, the assumption of a
uniform monochromatic wave may lead to misleadingly small detection
times. For this reason, we have developed a method for estimating the
detection time numerically for a more general waveform.

\subsection{An Improved Approach}

To better constrain the time required for detection of ripples in
distant clusters, we have developed a simple algorithmic approach to
determining whether a cluster profile contains ripples. The input to
the algorithm is a count rate profile either obtained from data or,
for our purposes in the current work, generated artificially according
to a method that will be described below. By generating artificial
count rate profiles, we will be able to determine how the detection
time for the ripples depends both on properties of the underlying
cluster atmosphere and properties of the ripples themselves.

The algorithm we use to determine whether a cluster has detectable
ripples is:

\begin{itemize}{}{}
\item Obtain a count rate profile $R$, either from data or from a model
\item Optionally (for simulated profiles) add Poisson noise to the
  count rate (assuming the input profile is an average count rate)
\item Convert the count rate profile to a surface brightness profile
  using the area of each annulus in the profile, to give an actual
  surface brightness profile $S$.
\item Fit a projected beta model to the surface brightness profile to
  give an underlying surface brightness profile $B$.
\item Calculate the fractional residuals of the actual surface
  brightness profile compared to the input surface brightness profile
  $a = (S-B) / B$
\item Fit a 1D smoothing spline to the surface brightness residuals
  $a$ to give a smooth profile $a_s$. The degree of smoothing of the
  spline was adjusted by eye to provide an acceptable compromise
  between over-smoothing the ripples and closely following the noise in $S$.
\item Use the zero points of $a_s$ to delimit peaks and troughs in the
  ripple.
\item For each peak or trough determine the significance of the excess
  or deficit in count rate in that region in the actual count rate $R$
  compared to the count rate that would result from the underlying surface
  brightness profile $B$.
\item Retain peaks or troughs only if their significance is greater
  than some threshold $\sigma$; typically $\sigma=3$, is chosen.
\end{itemize}

At the end of this process, we have a list of radii that mark the
end-points of regions containing significant excesses or decrements in
emission. However such regions do not necessarily constitute
ripples. In order that we see a region as a ripple there must be some
adjacent regions of positive or negative excess emission. The effect
of varying the number of adjacent regions required to constitute a
detection is examined in section \ref{sec:perseus_time}.

\subsection{Generating model counts profiles}

In order to use the procedure above to investigate the dependence of
detection time on the cluster and ripple properties, we require a
method for generating realistic count rate profiles for clusters
containing ripples. To generate these profiles, we start with
functional forms for the underlying surface brightness profile, $B$,
and a model for the modulus of the Fourier transform of the fractional
residuals (i.e. the square root of the power spectrum),
$|\mathscr{F}(R)|$. To generate the spatial surface-brightness
ripples, we assign a random phase to each frequency component in the
spectrum of the residuals, enforcing the condition that
$\mathscr{F}(R)(k) = \mathscr{F}(R)^{\dagger}(-k)$, so that $R(r)$ is
real. The resulting surface brightness profile $S = B(1+R)$ gives a
random wave with the desired power spectrum. We initially generate
such a surface brightness profile with a high spatial resolution and then
average over the radial bins appropriate to an observation to produce
the observed counts profile. Poisson noise appropriate to the
observation time is added when this surface brightness profile is
converted to a count rate. By using different random frequency
components, we are able to generate multiple spatial profiles and so
average the estimated detection times over multiple clusters with the
same ripple power spectrum.

\begin{figure}
  \includegraphics[width=\columnwidth]{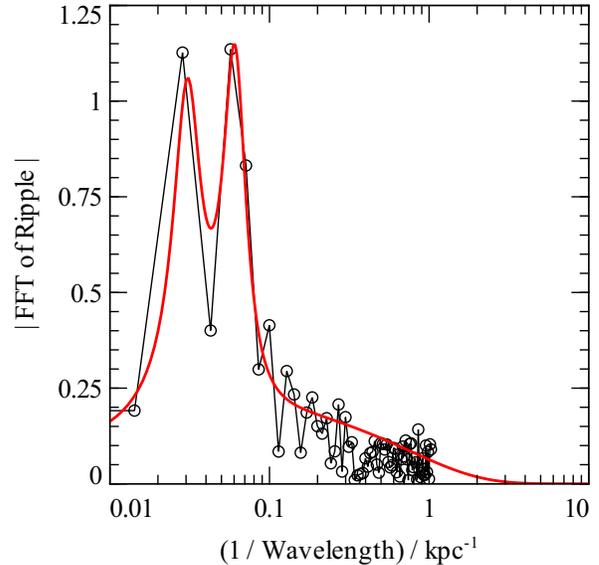}
  \caption{Positive frequency half of the Fourier transform of the
    fractional ripples in the Perseus cluster (points) and in
  our model (thick red line).}
\label{fig:power_spec_perseus}
\end{figure}

For our subsequent analysis, we often wish to compare results from
model clusters to those obtained for the Perseus cluster
data. Therefore we construct a fiducial model for the ripple power
spectrum which is based on the spectrum observed in the Perseus
cluster. We use the surface-brightness data shown in Fig. 3 of
\citet{Sanders2007}. This profile was generated in a sector with an
opening angle of approximately $0.13\times2\pi$, which we assume
throughout this paper is a typical angle over which ripples will be
coherent enough to allow a profile to be constructed from circular
annuli. The model spectrum along with the actual Perseus spectrum is
shown in Fig. \ref{fig:power_spec_perseus}. The model is of the form:

\begin{equation}
| \mathscr{F}(R(r)) |  = (1 - e^{-ak})*(L_0(k) + L_1(k) + B*e^{-ck})
\end{equation}
where $k$ is defined as $1/\lambda$ and $L_0$ and $L_1$ are Lorentzian
functions representing the main peaks in the surface brightness
profile:

\begin{equation}
L(k) = \frac{A\Gamma}{2\pi\left( (x-x_0)^{2} + \Gamma^{2}/4 \right)}
\end{equation}

Owing to the large number of free parameters in the model, the fit to
the Perseus model was performed by eye, aiming to recreate the main
features of the profile rather than provide an exact representation.

\subsection{Results}\label{sec:results}

\subsubsection{Observation Time to Detect Ripples in Perseus}\label{sec:perseus_time}

As a test of our method, we calculate the time required to detect
ripples in the Perseus cluster. The projected beta models were fitted
over the radius range $10-80\kpc$, and only ripples in this range were
counted. 

\begin{figure}
  \includegraphics[width=\columnwidth]{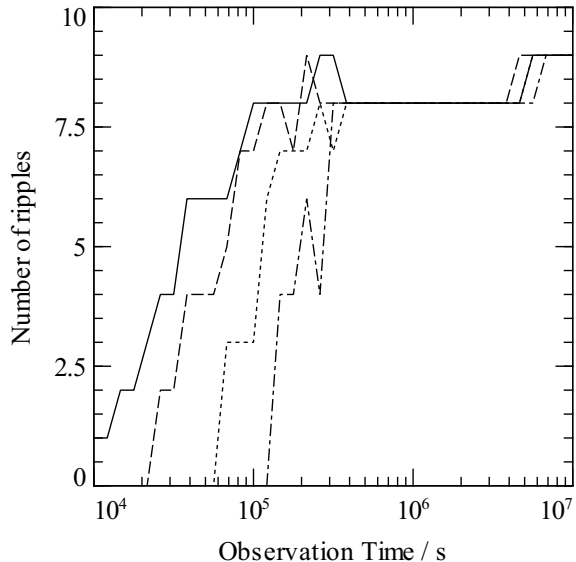}
  \caption{Number of ripples against observation time for the Perseus
    cluster with a detection requiring one (solid line), two (dashed
    line), three (dotted line) and four (dash-dotted) line features
    significant at the $3\sigma$ level.}
  \label{fig:perseus_number}
\end{figure}

Fig. \ref{fig:perseus_number} shows the expected number of ripples
detected against observing time for the Perseus cluster data. Each
line represents a different choice for the number of significant
excesses/decrements in emission needed to count a feature as a
ripple. To account for the random nature of the Poisson noise added to
our degraded surface brightness profiles, we have averaged the results
over several realisations of the counts profile. For the case where
2-3 adjacent features are needed to constitute a ripple, we start to
see several ripple-like features detected to $3\sigma$ significance in
about $40-100\ks$, broadly compatible with the bounds $ 20\ks < t <
200\ks$ imposed by the observed (non)detection times for ripples in
Perseus.
 
\begin{figure}
  \includegraphics[width=\columnwidth]{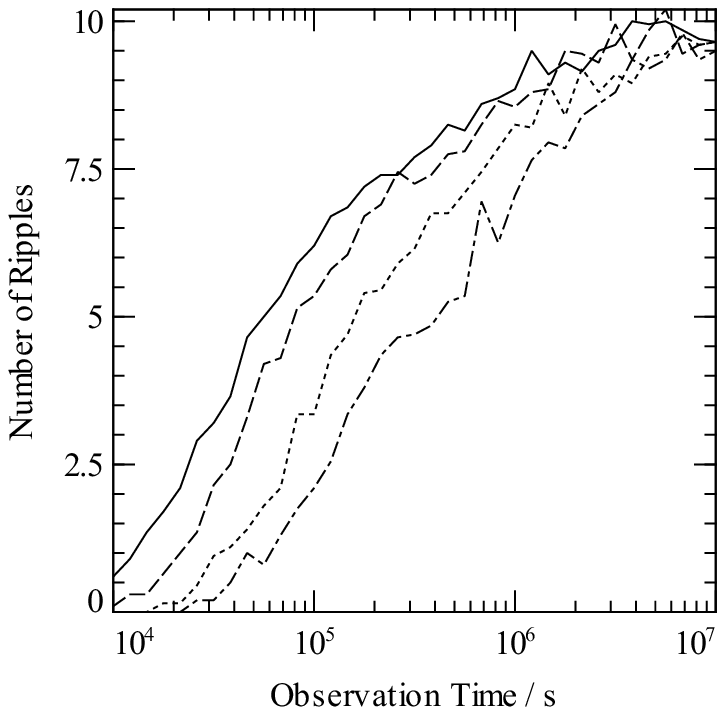}
  \caption{Number of ripples detected against observation time for an
    average over model clusters with ripples having the same power
    spectrum as the Perseus cluster, with a detection requiring one
    (solid line), two (dashed line), three(dotted line) and four
    (dash-dotted) line features significant at the $3\sigma$ level.}
  \label{fig:perseus_model_number}
\end{figure}

Fig. \ref{fig:perseus_model_number} shows the average number of
ripples detected against time for a set of model clusters with the
same ripple power-spectrum as Perseus. Again the different lines
represent a different number of adjacent features needed to constitute
a ripple. The number of ripples detected as a function of time is
comparable between the Perseus data and the model. In the case where a
single feature counts as a several ripples are detected in just over
$20\ks$, but in the more realistic case where $2-3$ adjacent features
are needed to identify a ripple detections take $40-100\ks$.

\subsubsection{Variation of Detection Time with Total Flux}

In order to determine how the time to detect a ripple varies with
various parameters of the underlying surface brightness and the ripple
structure, we use a binary search over the observation time. At each
observation time, several realisations of the model cluster are
constructed and we use the algorithm discussed above to search for
ripples in the region $r>10\kpc$. At each time, we
determine whether the median number of ripples detected is higher or
lower than some threshold. The time for the next step is chosen as
half time between the shortest time in which a detection has been made
and the longest time in which no detection has been made. Initially we
assume that ripples will always be detected in $10^7\s$ and will never
be detected in $10^4\s$ (violation of these assumptions will lead to
points lying just above the minimum or just below the maximum time,
which will be obvious in the results). We typically consider an
average of three ripple to be a detection and require at least two
adjacent significant features to define a ripple.

We use the same Perseus-like model cluster as before, with the same
radial bins. This is necessarily an approximation; in reality one
might choose larger bins for a fainter cluster. At each observation
time, 10 model clusters are generated to calculate the average number
of ripples.

\begin{figure}
  \includegraphics[width=\columnwidth]{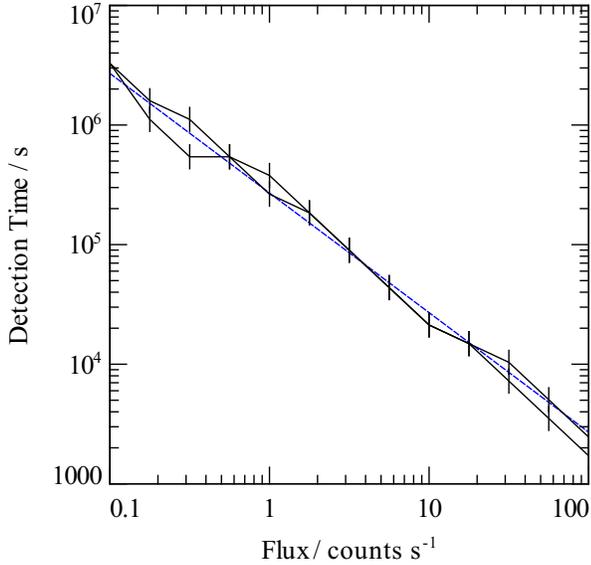}
  \caption{Two runs of the time required to detect Perseus-like
    ripples against the cluster flux in the detection region inside
    $100\kpc$. The dashed line shows $t\sim\text{flux}^{-1}$, as
    expected from equation \eqref{eqn:t_detect_analytic}, and is not a
    fit but is normalised for a detection time of $40\ks$ for the
    Perseus cluster (with a flux of $6.75 \ctps$ inside $100\kpc$ in
    the appropriate sector). The error bars account only for the accuracy of
    the binary search, not for the uncertainty in the time needed to
    make the detection. }
  \label{fig:time_flux}
\end{figure}

Fig. \ref{fig:time_flux} shows two independent
determinations of this variation in the required detection time with
the total flux of the cluster. The error bars represent the accuracy
at which the binary search is terminated, that is, the difference
between the maximum observation time for a non-detection and the
minimum for a detection. To gauge the uncertainty in the detection
time predicted by this method, we may simply compare points from the
independent runs; these seem to agree well, with the discrepancies
being similar in magnitude to the plotted error bars, indicating that
we are terminating the binary search at a sensible accuracy.

The detection time calculated for a Perseus-like cluster, with a flux
of $6.75$ counts per second inside $100\kpc$ is $\sim 40 \ks$.

\subsubsection{Variation in the Detection Time With Ripple Amplitude}
\begin{figure}
  \includegraphics[width=\columnwidth]{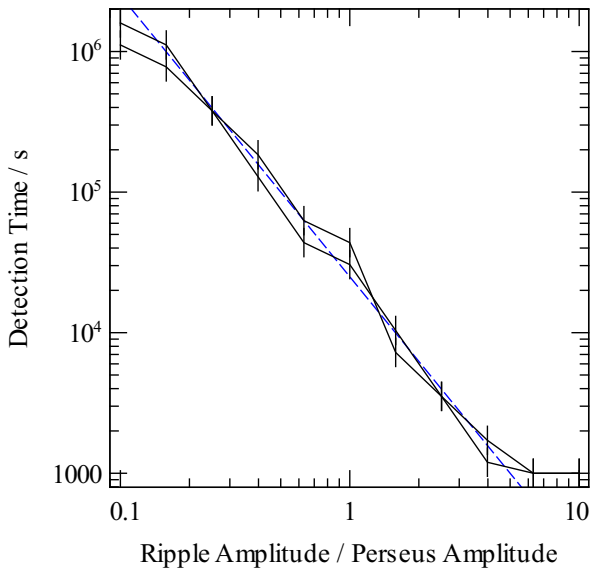}
  \caption{Two runs of the time required to detect Perseus-like
    ripples against the ripple amplitude. The dashed line shows $t\sim
    \text{amplitude}^{-2}$ scaled so the detection time at the Perseus
    ripple amplitude is $40\ks$.}
  \label{fig:time_amplitude}
\end{figure}

Fig. \ref{fig:time_amplitude} shows to independent calculations of the
detection time for various ripple amplitudes. The scaling of the
detection time with ripple amplitude is roughly $t\sim
\text{amplitude}^{-2}$, the same as lowest order in $f$ of equation
\eqref{eqn:t_detect_analytic}.

\subsubsection{Variation in Detection Time with Ripple Wavelength}

In order to determine the variation in detection time with the ripple
amplitude, we took the fiducial model and altered the values of $x_0$
in the two Lorentzian peaks, keeping the ratio of the two wavelengths
constant, whilst leaving the rest of the spectrum unaltered. This is
inevitably an over-simplification; the underlying ripple wavelength is
determined by the inflation process, which will also affect the other
frequency components. Also, as discussed in section
\ref{sec:projection}, a ripple of smaller wavelength and equal
amplitude in projection corresponds to a larger amplitude ripple in
3D.

\begin{figure}
  \includegraphics[width=\columnwidth]{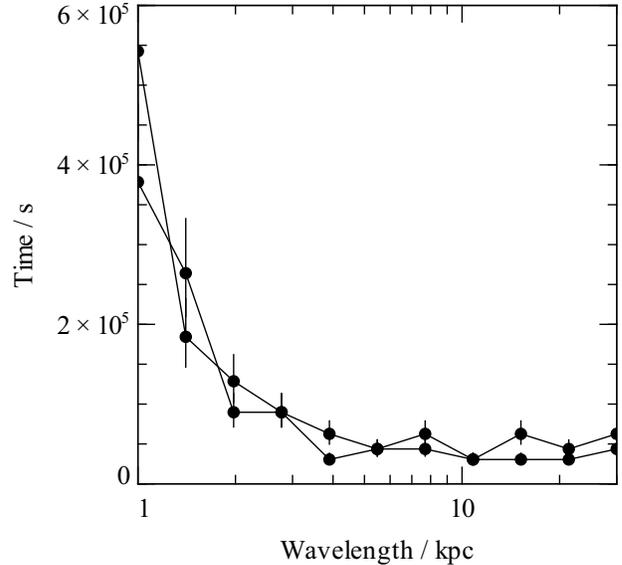}
  \caption{Two runs of the time required to detect Perseus-like
    ripples against the ripple wavelength. The wavelength shown
    corresponds to the highest frequency Lorentzian peak in the model.}
  \label{fig:time_wavelength}
\end{figure}

\begin{figure}
  \includegraphics[width=\columnwidth]{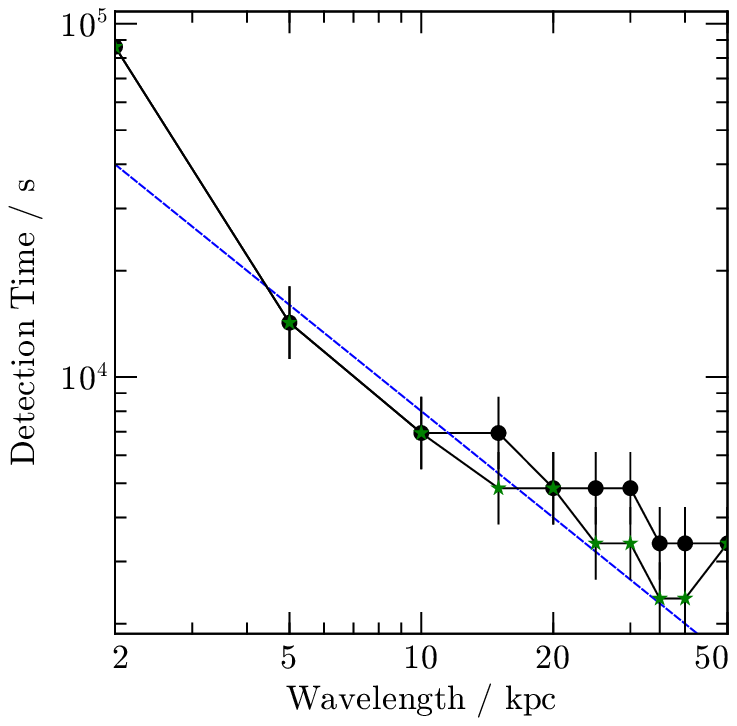}
  \caption{Time to detect monochromatic ripples against the ripple
    wavelength for underlying surface brightness profiles of beta
    model form with $r_\text{core}=10\kpc$ (circles) and with a
    power-law profile (green stars). The profiles are scaled so that
    the detection times at $r=10\kpc$ are similar. The blue dashed
    line indicates $t\sim \lambda^{-1}$.}
  \label{fig:time_wavelength_mono}
\end{figure}

Fig. \ref{fig:time_wavelength} shows the variation in detection time
with the ripple wavelength. The detection time appears to be almost
independent of wavelength above wavelengths of about $8\kpc$. This
result is inconsistent with a naive prediction; the number of counts
in the peak of a ripple is $N \sim SA$ where $S$ is the surface
brightness at the ripple radius and $A$ is the area covered by the
ripple. Typically, we have $S\sim r^{-1}$ so $N \sim r^{-1} r\lambda
\sim \lambda$, and so we expect the time to detect the ripple scales like
$\sim1/\lambda$. To understand this discrepancy, we have run
simulations using a simple monochromatic ripple profile, shown in
Fig. \ref{fig:time_wavelength_mono}. With monochromatic waves, the
detection time more closely follows $t\sim1/\lambda$ out to large
ripple wavelengths. There is also some evidence that the beta-model
underlying profile has larger detection times at radii above $r\sim
r_\text{core}$ than the scale-free power-law model.

\subsubsection{Variation in Detection Time with Underlying Cluster Properties}

\begin{figure}
  \includegraphics[width=\columnwidth]{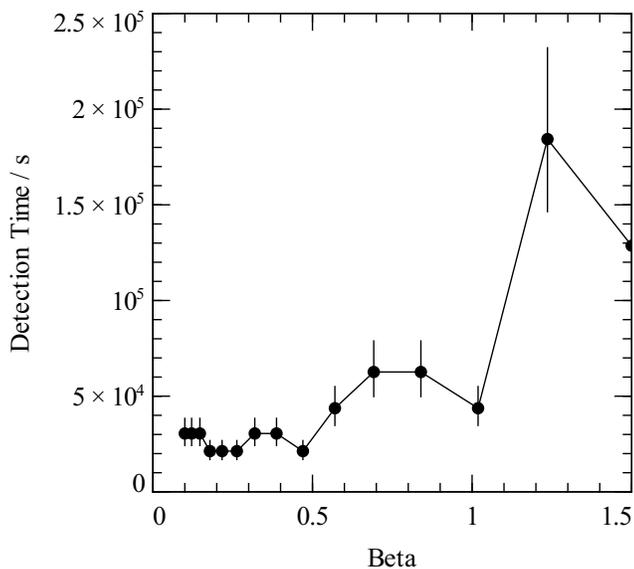}
  \caption{Mean time for detection against underlying cluster beta parameter.}
  \label{fig:time_beta}
\end{figure}

Figure \ref{fig:time_beta} shows the variation in detection time with
the $\beta$ parameter of the underlying beta model. For reasonable
values of $\beta$ there is little evidence of a substantial change in
the detection time, although the time becomes large for very large
$\beta$ values. Variations in the core radius of the beta model
similarly make little difference to the results.


\subsubsection{Variation in Detection Time With Redshift}

\begin{figure}
  \includegraphics[width=\columnwidth]{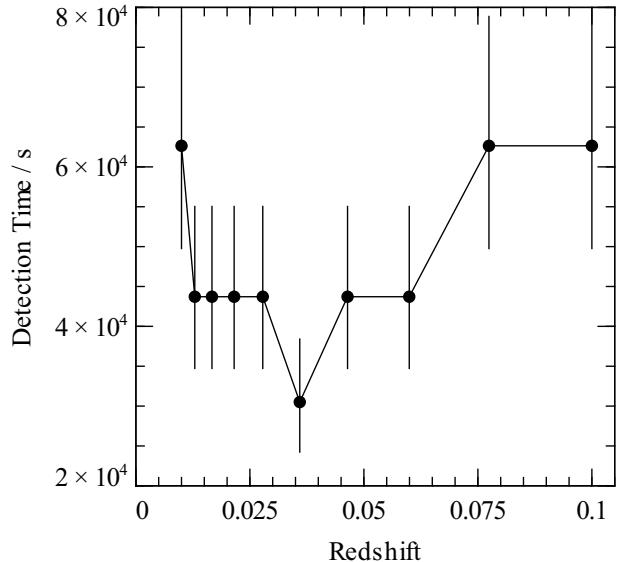}
  \caption{Mean time for detection against cluster redshift, assuming
    the cluster flux is constant, independent of redshift.}
  \label{fig:time_redshift}
\end{figure}

Fig. \ref{fig:time_redshift} shows the variation of detection time
with redshift at constant flux. This means that redshift variations
correspond to a rescaling of the cluster's physical dimensions. For the
cluster parameters and range of redshifts shown, the detection time
does not vary significantly, however there will be a point where the
wavelength of the ripple is comparable to the size of the angular bins
at the cluster redshift. After this point, which will occur at smaller
redshift for smaller wavelength, the ripples will become significantly
more difficult to detect.

\section{Discussion}

\subsection{Time Needed to Detect Ripples}

Assuming that the different parameters in section \ref{sec:results}
may be treated independently, we may derive a simple analytical
approximation for the time to detect ripples in a given cluster. If we
assume that the ripples are long enough wavelength that the scaling of
detection time with ripple wavelength may be ignored, the only strong
dependencies seen are with flux and amplitude, which scale in a manner
consistent with the simple analytical model; $t \sim \text{flux}^{-1}$
and $t \sim \text{amplitude}^{-2}$. Taking the normalisation from the
Perseus model:
\begin{equation}
  t = 40\left(\frac{f}{f_\text{Perseus}}\right)^{-1} 
  \left(\frac{a}{a_\text{Perseus}}\right)^{-2}\ks\label{eqn:detect_time}
\end{equation}
for a cluster of flux $f$ and ripple amplitude $a$. Except for the
wavelength scaling, this result is very close to the analytic
prediction of \eqref{eqn:t_detect_analytic}, with the normalization
increased by a factor $\sim3.5$. 

\subsection{Detectability of sound waves in nearby clusters}\label{sec:localclusters}

To investigate the feasibility of detecting ripples in clusters other
than Perseus using the current generation of X-ray satellites, we
have used equation \eqref{eqn:detect_time} to determine the regions of
$a/a_\text{Perseus}$, $f/f_\text{Perseus}$ parameter space accessible
to \chandra\ observations of different lengths. We have then estimated
appropriate values for $a/a_\text{Perseus}$ and $f/f_\text{Perseus}$
for several nearby clusters.

To estimate $f/f_\text{Perseus}$, we need the flux in $\ctps$ over the
region where ripples could be detected in the cluster. We assume this
corresponds to the region $2r_\text{bubble centre} - 5_\text{bubble
  centre}$ , where $r_\text{bubble centre}$ is the average radius of
the bubble from the centre of the cluster taken from \citet{Dunn2004}
and \citet{Dunn2005}. Making different assumptions about where the
flux should be measured does not change our results substantially.  To
measure the flux, we use raw \chandra\ events files taken from
observations in the \chandra\ archive. In all cases except Virgo, the
observations are taken with the ACIS-S instrument; for Virgo we scale
from ACIS-I to ASIS-S by increasing the count rate by a factor 1.5
appropriate for a $3\kev$ plasma. The energy range is limited to
$0.5-7\kev$. In each case, obvious point sources
in the region of interest were removed by eye and a blank-sky
background scaled to the correct exposure time and area was
subtracted.

To estimate the amplitude of the ripples, we use the relationship
between the power in a sound wave and its pressure amplitude
\cite{LandauLifshitzFluids}:
\begin{equation}
P_\text{wave} = 4\pi r^2\frac{(\delta p)^2}{\rho c}
\end{equation}
Assuming that the effect of projection is to reduce the projection
reduces the surface brightness perturbation so $\delta S/S = \kappa
\delta n/n$ -- as discussed in Section \ref{sec:projection} -- the
amplitude of surface brightness fluctuations is related to the power
in the wave as:
\begin{equation}\label{eqn:delta_sb}
\frac{\delta S}{S} = \frac{\kappa
  P_\text{wave}^{\frac{1}{2}}}{\sqrt{4\pi\gamma^{3/2}}r}\left(
  \frac{\rho}{p^3} \right)^\frac{1}{4}
\end{equation}

To estimate the wave power in this expression, we make use of two
approaches; an estimate that $P_\text{wave} = P_\text{cavity}$ based
on the cavity power using the buoyancy timescale \cite{Birzan2004,
  Dunn2004, Rafferty2006} and an estimate $P_\text{wave} =
L_\text{cool}(1-2r_\text{bubble}/r_\text{cool})$. The factor
$(1-2r_\text{cool}/r_\text{bubble})$ is to account for the fact that
the very central region of the cluster is not likely to be heated by
sound waves but instead by weak shocks and cavity heating
\citep{McNamaraNulsen2007}.  To estimate the projection factor
$\kappa$, we assume that the dominant ripple wavelength is $\sim 2
r_\text{b}$ i.e. twice the average bubble dimension; this is a good
approximation for the Perseus cluster. $\kappa$ is then calculated
using the procedure in section \ref{sec:projection}. We take $r$ to be
$5r_\text{bubble centre}$.

\begin{figure}
  \includegraphics[width=\columnwidth]{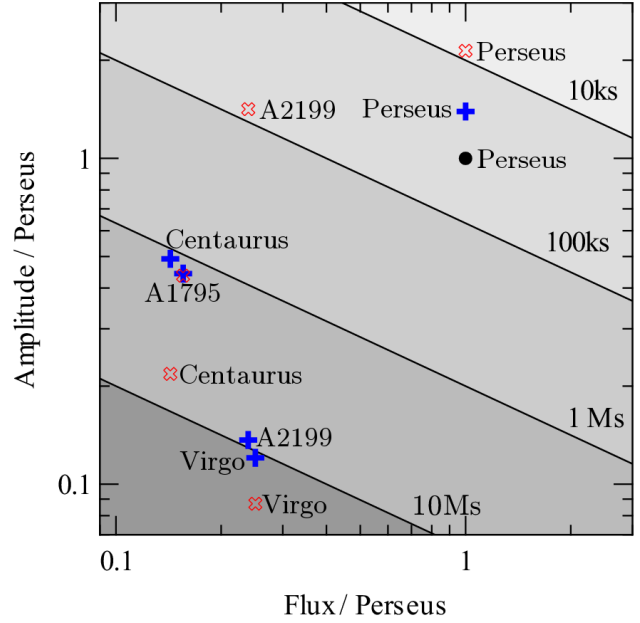}
  \caption{Flux and predicted ripple amplitude of some local clusters
    compared to Perseus. Blue plusses are based on the X-ray
    luminosity within the cluster cooling radius, red crosses are an
    estimate based on the cavity power. The black circle indicates the
    observed position of the Perseus cluster. Shaded regions indicate the
    predicted detection times for sound waves in the clusters using
    \chandra .}
  \label{fig:cluster_times}
\end{figure}

Using this model, with the cluster and bubble properties presented in
\citet{Dunn2004, Dunn2006} assuming that the cavity power is $4pV$
appropriate if the bubbles are filled with a fully-relativistic
plasma, we predict the value of $\delta S/S$ for several local
clusters, which we compare to $\delta S / S = 0.05$ for the Perseus
cluster.  Fig. \ref{fig:cluster_times} shows the position of these
clusters on a flux-amplitude plane, including values for the Perseus
cluster both calculated from this model and from the actual
observations. The corresponding detection times with \chandra\ are
indicated by the shaded regions.

The above analysis suggests that detecting ripples in several local
clusters may be possible with the \chandra\ satellite. In particular
Centaurus, Abell 1795 and Abell 2199 are promising candidates
for detections in around $1\ms$ of observation time. However it is
important to note that there is considerable uncertainty in the wave
amplitude determined for each cluster, with the two calculated
estimates giving up to an order of magnitude difference in the
estimated amplitude. In Perseus it is apparent that the cooling
luminosity is a better estimator of the wave amplitude than the cavity
power, which is likely true for other clusters if the power in sound
wave is closer to the time-average heating power than the
instantaneous cavity power. 

In addition to the power estimates, there are also uncertainties
associated with the ripple wavelength and the radius at which it is
assumed that the detection will be made. For a cluster like Centaurus
which is currently under-heated by the cavities, the final size of the
bubbles is likely greater than the current size; \citet{Dunn2006}
estimate $r_\text{b}/r_\text{b, max} = 0.75$ for Centaurus, although
it is notable that many clusters in their sample have
$r_\text{b}/r_\text{b, max} > 1$, indicating the cavities grow larger
than needed to offset the cooling. Increasing the bubble radius will
increase the wavelength of the ripples generated by the cavity
inflation, reducing the projection factor for the ripples. For
Centaurus, increasing the wavelength by a factor 1.5 would reduce the
time needed to detect the waves by a factor of $\sim2.25$ to a few
hundred kilo-seconds. Conversely, the most prominent bubbles in Abell
2199 are likely detached from the central source and buoyantly rising
in the ICM. Therefore their dimensions likely
overestimate the ripple wavelength and excluding the region inside
$2r_\text{bubble centre}$ probably underestimates the cooling
luminosity. This may explain the large discrepancy between the cooling
luminosity and cavity luminosity based estimates of the amplitude in
this system.

Detection times may also increase if the cluster is in a regime where
the wavelength of the ripples is important; for example if it has
astrongly monochromatic ripples so the detection time scales as $\sim
1/\lambda$.

Given the large uncertainty on the amplitude of the waves, and the
strong dependence of the detection time on the amplitude, it may be
possible to see ripples in some local clusters with currently
available data. In Abell 2199, an isothermal shock has been seen in a
$33\ks$ exposure, but an unsharp-mask analysis revealed no further
ripple-like features. For Centaurus, $200\ks$ of \chandra\ data are
available, and have been analysed for ripples by Sanders et al. (in prep.).

It is also important to recognise the limitations of our predictions
compared to the procedure that has been used to find ripples in
practice. The starting place to locate ripples in practice will be an
image of the cluster such as Fig. \ref{fig:per_ripples} in which the
small-scale features have been brought out using a technique such as
unsharp-masking or a Fourier-filter. Assuming ripples exist in the
cluster, whether they are seen in the resulting image will depend
substantially on the degree of coherence of the ripples and the
complexity of the radial power spectrum of the ripples. Ripples with a
high degree of coherence might be discovered in much less time than
suggested by our analysis whilst those with a limited degree of
coherence might not be noticed even in very long exposures. Since
coherent features are likely easier to identify with better spatial
resolution this suggests ripples will be easier to detect in
low-redshift clusters than in higher redshift clusters even where
their observed flux is similar.

Related to this issue is that of the threshold number of standard
deviations from the background count rate needed to class a feature as
a ripple. In our analysis, we have assumed each feature needs to be
detected to $3\sigma$. In practice, detecting a large number of
clearly ripple-like features to less than $3\sigma$ may be a more
convincing detection than a detection in which a few blobs are
detected to higher significant, but the intermediate structure is not
clearly ripple-like. 

Given these limitations, extreme caution must be made in predicting
exact detection times for ripples and for interpreting non-detections
as indicating that no ripple-like structures exist in a given cluster.

\section{Conclusion}

In order to understand the role played by sound waves in distributing
energy in cluster cores, it is essential to study these waves in a
variety of cluster systems. To understand the requirements for such a
study, we have investigated the effect of projection on reducing the
observed wave amplitude compared to the intrinsic amplitude in a range
of cluster atmospheres, and we have calculated the detection times for
waves in a number of cluster environments.

Projection of ripples in the emissivity profile substantially reduces
the amplitude of the surface brightness ripples. The magnitude of this
effect is a strong function of the ripple wavelength, but depends
little on the underlying atmosphere properties. If the wavelength of
the ripples is correlated with the dimensions of the cavity, this
implies systems with larger cavities may be more promising targets for
the detection of ripples.

By constructing an algorithm for detecting ripples in model data, we
have shown that the detection time for Perseus-like surface
brightness ripples is critically determined by two features of the
cluster -- the total flux and the ripple amplitude. Other factors such
as the underlying cluster properties have a much more limited
effect. Applying our results to nearby clusters, and assuming that the
ripple amplitude is sufficient to heat the cluster, we estimated
detection times for ripples using the \chandra\ satellite.

These detection times suggest that a selection of nearby, bright
clusters may contain ripples detectable in around $\sim1\ms$ of
\chandra\ observation time, although there is considerable uncertainty
brought about by uncertainties in the expected amplitude of the
ripples. In cooler clusters such as Virgo that require less heating,
and so are expected to have smaller ripples, the detection times are
likely prohibitively long with \chandra\ but should be well within
reach of \xeus\ which promises one to two orders of magnitude greater
effective area.

\section{Acknowledgements}
ACF thanks the Royal Society for support.

\bibliographystyle{mnras} 
\bibliography{refs}

\begin{thebibliography}{}

\bibitem[\protect\citeauthoryear{{B{\^i}rzan} et~al.}{{B{\^i}rzan}
  et~al.}{2004}]{Birzan2004}
{B{\^i}rzan} L., {Rafferty} D.~A., {McNamara} B.~R., {Wise} M.~W.,  {Nulsen}
  P.~E.~J., 2004, \apj, 607, 800

\bibitem[\protect\citeauthoryear{{B\"{o}hringer} et~al.}{{B\"{o}hringer}
  et~al.}{1993}]{Bohringer1993}
{B\"{o}hringer} H., {Voges} W., {Fabian} A.~C., {Edge} A.~C.,  {Neumann} D.~M.,
  1993, \mnras, 264, L25

\bibitem[\protect\citeauthoryear{{Cavaliere} \& {Fusco-Femiano}}{{Cavaliere} \&
  {Fusco-Femiano}}{1976}]{Cavaliere1976}
{Cavaliere} A.,  {Fusco-Femiano} R., 1976, \aap, 49, 137

\bibitem[\protect\citeauthoryear{{Dunn} \& {Fabian}}{{Dunn} \&
  {Fabian}}{2004}]{Dunn2004}
{Dunn} R.~J.~H.,  {Fabian} A.~C., 2004, \mnras, 355, 862

\bibitem[\protect\citeauthoryear{{Dunn} \& {Fabian}}{{Dunn} \&
  {Fabian}}{2006}]{Dunn2006}
{Dunn} R.~J.~H.,  {Fabian} A.~C., 2006, \mnras, 373, 959

\bibitem[\protect\citeauthoryear{{Dunn}, {Fabian}, \& {Taylor}}{{Dunn}
  et~al.}{2005}]{Dunn2005}
{Dunn} R.~J.~H., {Fabian} A.~C.,  {Taylor} G.~B., 2005, \mnras, 364, 1343

\bibitem[\protect\citeauthoryear{{Ettori}}{{Ettori}}{2000}]{Ettori2000}
{Ettori} S., 2000, \mnras, 318, 1041

\bibitem[\protect\citeauthoryear{{Fabian} et~al.}{{Fabian}
  et~al.}{2003}]{Fabian2003}
{Fabian} A.~C., {Sanders} J.~S., {Allen} S.~W., {Crawford} C.~S., {Iwasawa} K.,
  {Johnstone} R.~M., {Schmidt} R.~W.,  {Taylor} G.~B., 2003, \mnras, 344, L43

\bibitem[\protect\citeauthoryear{{Fabian} et~al.}{{Fabian}
  et~al.}{2006}]{Fabian2006}
{Fabian} A.~C., {Sanders} J.~S., {Taylor} G.~B., {Allen} S.~W., {Crawford}
  C.~S., {Johnstone} R.~M.,  {Iwasawa} K., 2006, \mnras, 366, 417

\bibitem[\protect\citeauthoryear{{Landau} \& {Lifshitz}}{{Landau} \&
  {Lifshitz}}{1959}]{LandauLifshitzFluids}
{Landau} L.~D.,  {Lifshitz} E.~M., 1959, {Fluid mechanics}.
\newblock Course of theoretical physics, Oxford: Pergamon Press, 1959

\bibitem[\protect\citeauthoryear{{McNamara} \& {Nulsen}}{{McNamara} \&
  {Nulsen}}{2007}]{McNamaraNulsen2007}
{McNamara} B.~R.,  {Nulsen} P.~E.~J., 2007, \araa, 45, 117

\bibitem[\protect\citeauthoryear{{Peterson} \& {Fabian}}{{Peterson} \&
  {Fabian}}{2006}]{PetersonFabian2006}
{Peterson} J.~R.,  {Fabian} A.~C., 2006, \physrep, 427, 1

\bibitem[\protect\citeauthoryear{{Peterson} et~al.}{{Peterson}
  et~al.}{2003}]{Peterson2003}
{Peterson} J.~R., {Kahn} S.~M., {Paerels} F.~B.~S., {Kaastra} J.~S., {Tamura}
  T., {Bleeker} J.~A.~M., {Ferrigno} C.,  {Jernigan} J.~G., 2003, \apj, 590,
  207

\bibitem[\protect\citeauthoryear{{Rafferty} et~al.}{{Rafferty}
  et~al.}{2006}]{Rafferty2006}
{Rafferty} D.~A., {McNamara} B.~R., {Nulsen} P.~E.~J.,  {Wise} M.~W., 2006,
  \apj, 652, 216

\bibitem[\protect\citeauthoryear{{Sanders} \& {Fabian}}{{Sanders} \&
  {Fabian}}{2007}]{Sanders2007}
{Sanders} J.~S.,  {Fabian} A.~C., 2007, \mnras, 381, 1381

\end{thebibliography}

\end{document}